\documentclass[aps,pra,twocolumn,noshowpacs,preprintnumbers,amsmath,amssymb,footinbib]{revtex4}
\usepackage{graphicx,epsfig}
\usepackage{bm}
\usepackage{dcolumn}
\usepackage{xcolor}
\usepackage[breaklinks=true,colorlinks,citecolor=blue,linkcolor=blue,urlcolor=blue]{hyperref}
\usepackage{braket}
\usepackage{mathrsfs}
\usepackage{graphicx,graphics}
\begin{document}

\title{High-dimensional states of light with full control of transverse path and OAM degrees of freedom}

\author{D.\ Pab\'on$^{1,2,*}$}\email{dudbilopr@gmail.com}
\author{S. Ledesma$^{1,2}$}
\author{L.\ Reb\'on$^{3,4}$}
\affiliation{$^{1}$Departamento de F\'isica, FCEyN, Universidad de Buenos Aires, Pabell\'on 1, Ciudad Universitaria,  Buenos Aires (1428), Argentina.\\ $^{2}$Consejo Nacional de Investigaciones Cient\'ificas y T\'ecnicas (CONICET), Argentina.\\ $^{3}$Departamento de F\'{\i}sica, FCE, Universidad Nacional de La Plata, C.C. 67, La Plata (1900), Argentina. \\ $^{4}$Instituto de F\'isica de La Plata, UNLP - CONICET, Argentina.}
\date{\today}

\begin{abstract}
We present here a compact scheme for the generation of high-dimensional states of light encoded in the transverse path variable of photons that carry orbital angular momentum. We use a programmable spatial light modulator in phase configuration to create correlations between these two spatial degrees of freedom. With our setup we are able to control, independently, the relative phases and amplitudes of the path superposition in addition to the topological charge of each path. Moreover, we engineer correlations that emulate bipartite quantum states of dimensions $d\times m$. Experimental results from the characterization of different generated states of dimensions up to $9\times 5$ are in excellent agreement with the numerical simulations. Fidelity with the target state is, for all cases, above $95\%$. 
\end{abstract}


\maketitle
It is well known that photons can encode information in several degrees of freedom (DoFs) as polarization~\cite{PhysRevLett754337}, frequency~\cite{Lukens2017} or time-bin modes~\cite{Brecht2015}.
Of particular interest, in a scenario of quantum communications, is the use of the transverse spatial modes of light with orbital angular momentum (OAM). 
Photons carrying OAM~\cite{Allen92} have been used to encode $d$-dimensional quantum states, called {\it qudits}, allowing to increase the quantum complexity
without increasing the number of involved particles~\cite{Erhard2018}. 

In this context, OAM-based communications could achieve a higher data transmission rate through free space~\cite{Krenn13648} or optical fibers~\cite{Zhu2018}, and improve security against a potential eavesdropping~\cite{Mirhosseini_2015}.  
More recently, different DoFs of multipartite-photonic states have been combined to increase the dimension of the system and its quantum complexity ~\cite{He2017,Wang2018}.
Additionally, the linear superposition of multiple-DoFs states of a single photon, or even the use of various correlated DoFs of classical light~\cite{Rafsanjani2015,Aiello2015}, provides access to larger systems and enables the generation of non-separable states, 
useful for some applications of quantum information~\cite{ndagano2017characterizing}
and optical metrology~\cite{Tppel2014}.

The experimental generation of light with an OAM spectrum can be done using phase or amplitude masks~\cite{RosalesGuzman2017}. One of the techniques to accomplish this task consists of displaying a fork hologram in a programmable device, like a liquid crystal display (LCD), forming part of a spatial light modulator (SLM) system. These SLMs have also proven useful for generating qudits codified in the transverse spatial DoF of light with $\ell=0$, by means of the discretization of the transverse linear momentum of single photons~\cite{Lima2009}.
In this case, the encoding process is achieved when photons pass through an aperture with $d$ slits (alternatively, through $d$ bi-dimensional spatial regions) and the dimension of the state is defined by the amount $d$ of possible paths followed by a photon. As it was demonstrated~\cite{SolsProsser2013PreparingAP,Varga2014}, the complex superposition that defines the transverse-path qudit can be obtained on a single phase-only SLM system that controls the phase of each slit and encodes its real amplitude. 

In this letter we present a novel architecture that exploits the versatility of programmable SLMs to correlate two spatial DoFs of light: the transverse path and the OAM modes. Our method consists of programming a phase hologram in each of the $d$ regions of a LCD screen, generating $d$ orthogonal transverse-path states that carry a given $\ell$ quanta of OAM each. By using a SLM operating in a phase only mode we are able to control both the OAM contribution and the complex amplitude of each path, regardless of the OAM content. As a result, we can obtain an arbitrary superposition of the OAM-transverse path states giving a non-separable state of light. Besides, we exploit the regions of the LCD that are not used to encode information to display a blazed grating, whose diffracted order, is used as a reference beam in a later characterization step. In this way, we have implemented a new compact design where only one SLM generates states of light that emulate bipartite quantum systems of dimension $d\times m$, and the reference for the  characterization of the state carried out by means of a phase shifting interferometry (PSI) technique~\cite{Creath1988}. 

\textit{Central issues.}
We will start by describing the generation of
OAM-transverse path states. Let us suppose that $\mathcal{U}(\boldsymbol{\rho},z)$ denotes a paraxial and monochromatic light field,
whose transverse
spatial profile is proportional to the transverse probability amplitude of a single-photon field.
The variable ${\boldsymbol{\rho}}=\rho~(\hat{{\boldsymbol{x}}}\cos{\theta}+\hat{{\boldsymbol{y}}}~\sin{\theta})$ represents the transverse coordinates and the $z$-axis is taken along the direction of propagation. At $z=0$ this field passes through an aperture described by the complex transmission function $T(\boldsymbol{\rho})$, so that it is transformed from $\mathcal{U}(\boldsymbol{\rho})\equiv\mathcal{U}(\boldsymbol{\rho},z=0)$  to  $~\widetilde{\mathcal{U}}(\boldsymbol{\rho})\equiv T(\boldsymbol{\rho})~\mathcal{U}(\boldsymbol{\rho})$.
We will consider, without loss of generality, a transmittance defining an array of $d$ circular spatial regions of radius $r$, $T(\boldsymbol{\rho})=
\sum_{\mu=0}^{d-1}\text{circ}\left(\frac{\mid\boldsymbol{\rho}-\boldsymbol{\rho}_\mu\mid}{r}\right)~f_{\mu}(\boldsymbol{\rho}-\boldsymbol{\rho}_\mu)$.
Here $\text{circ}(\rho) = 1$ if $\rho\leq 1$, zero otherwise, and $f_\mu(\boldsymbol{\rho}-\boldsymbol{\rho}_\mu)$ is the complex modulation function introduced by the SLM in the $\mu$-region, corresponding, in our implementation, to a diffraction grating with fork dislocations centered on $\boldsymbol{\rho}_\mu$.
Under the condition  $\mid\boldsymbol{\rho}_\mu-\boldsymbol{\rho}_{\mu'}\mid~>2r~~\forall \mu,\mu'$, such an aperture divides a Gaussian wavefront in $d$ non-overlapping vortex beams with topological charge $\ell_\mu$. Therefore, the transformed field can be written as $\widetilde{\mathcal{U}}(\boldsymbol{\rho})=\sum_{\mu=0}^{d-1}\beta_\mu~\mathcal{U}_{\ell_\mu}(\boldsymbol{\rho}-\boldsymbol{\rho}_\mu)~e^{~i\ell_\mu\theta_\mu}$,
where $e^{~i\ell_\mu\theta_\mu}$ is an OAM eigenstate with $\ell_{\mu}\hbar$ of OAM per photon in the electromagnetic field, and $\mathcal{U}_{\ell_\mu}(\boldsymbol{\rho})\equiv\text{circ}(\rho/r)~A_{\ell_\mu}(\rho)$ is the radial amplitude function of this mode, which in general will depend on $\ell_\mu$. 
The optical vortex in each transverse path is defined with respect to the individual center $\boldsymbol{\rho}_\mu$ so the angular coordinate is $\theta_\mu= \text{arg}(\boldsymbol{\rho}-\boldsymbol{\rho}_\mu)$.
It should be note that the complex amplitude $\widetilde{\mathcal{U}}(\boldsymbol{\rho})$ of the resulting $d$-fold beams at $z=0$ is described by a non-separable function given by the correlations between two independent DoFs. It emulates a bipartite state, which in standard notation for quantum states, can be written as
\begin{equation}
    \ket{\psi} 
     = \sum_{\mu=0}^{d-1}{\tilde{\beta_{\mu}}\ket{\mu}\otimes\ket{\ell_{\mu}}},\label{eq1}
\end{equation}
with complex coefficients $\tilde{\beta_{\mu}}=\beta_\mu/\sqrt{\sum_{i=0}^{d-1}\mid\beta_{i}\mid^2}$ and kets $\ket{\mu}$, $\ket{\ell_{\mu}}$ representing the transverse-path state and the OAM state, respectively. The dimension of the state is given by the number of transverse paths $d$, and the cardinality of the finite set of OAM used in the codification, $m=\mathcal{C}(\{\ell_0,\ell_1,...,\ell_{d-1}\})$.
These two DoFs are independent in the sense that it is possible to set the amount of OAM per photon in each of the $d$ available transverse paths. In addition, we can control the complex coefficients $\beta_\mu$ that define the superposition, i. e., the intensity and the relative phase of each transverse path.

\begin{figure}[h]
\centering
\includegraphics[width=0.45\textwidth]{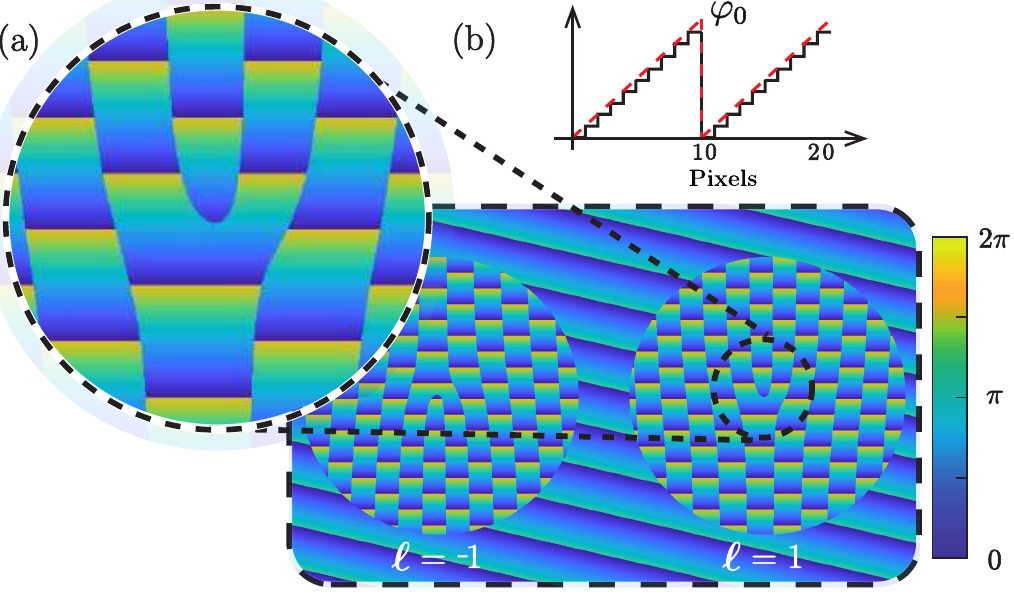}
\caption{(a) Phase mask representing the state $\ket{\psi_1}=\frac{\ket{0-1}+\ket{11}}{\sqrt{2}}$. The state is codified in the circular regions while the background generates a reference beam used in the state reconstruction stage. (b) Schematic profile of an ideal (red) and real (black) blazed grating.}\label{fig_1}
\end{figure}

To prepare the state in Eq.~(\ref{eq1}), the LCD display is divided in $d$ circular regions showing a 
phase hologram each. These holograms are fork-shaped containing the OAM information given by the number of central dislocations. When such a hologram with $\ell$ dislocations is illuminated with a plane wave, two beams are generated with a topological charge $\pm\ell$ and a doughnut-shaped intensity in the positive and negative first diffraction order, respectively~\cite{Pabon20172}. With the purpose of controlling the complex amplitude $\tilde{\beta_{\mu}}$ and to redirect the information in the first diffraction order for a spatial filtering, each hologram is combined with a blazed grating. This codification is achieved in a single phase-only SLM by adopting a similar procedure to that described in Refs.~\cite{SolsProsser2013PreparingAP,Varga2014}. For the present design, the following consideration must 
be done:

i. A binary $0-\pi$ phase hologram multiplied by a phase diffraction blazed grating is displayed in the spatial region defining a particular transverse-path state. This design allows the first order to be displaced not only in the horizontal direction $(\hat{{\boldsymbol{x}}})$, but also in the vertical direction $(\hat{{\boldsymbol{y}}})$ in the far field. 
Thus, the first order can be easily filtered and the
light distribution is not corrupted by unwanted noise.
As an example, in Fig.~\ref{fig_1}(a) we show the phase mask corresponding to the state $\ket{\psi_1}=\frac{1}{\sqrt{2}}\left(\ket{\mu=0}\otimes\ket{\ell=-1}+\ket{\mu=1}\otimes\ket{\ell=1}\right)$.
The	typical single-fork dislocation corresponding to $\ell=1$ is clearly seen in the zoomed image.

ii. The efficiency of the first diffracted order for an ideal blazed grating can be calculated by the expression $\eta_1 = \text{sinc}^2\left(1-\frac{\varphi_0}{2\pi}\right)$, where $\varphi_0$ is the phase modulation depth of the grating. In a real grating there is a discretization in the phase levels of the blazed profile given by the number of pixels $N$ in a grating period~\cite{goodman2005introduction}. A schematic profile of a blazed grating with $N=10$ is shown in Fig.~\ref{fig_1}(b). Therefore, it is possible to modulate the amount of light diffracted by each region by selecting the phase modulation $\varphi_0$, and consequently, the weights in the superposition of Eq.~(\ref{eq1}). We assigned the maximum value of the real amplitude $|\tilde{\beta}_{\mu}|=1$ to the maximum grating efficiency reached with our SLM.

iii. A lateral displacement of the grating by $\Gamma$ pixels corresponds, in the far field, to an additional constant phase in the $n$-th diffraction order of $\delta=2\pi \frac{\Gamma}{p}n$, where $p$ is the grating period measured in pixel units. Hence, introducing relative displacements between blazed gratings of different regions, we set the value $\text{arg}(\tilde{\beta}_{\mu})$. In this way, we have a complete control of the complex coefficients $\tilde{\beta}_{\mu}$ in equation Eq.~(\ref{eq1}). Finally, after a spatial filtering, we obtain the desired OAM-transverse path state in the first diffraction order.

We have implemented a four-step PSI method~\cite{Creath1988} to characterize the generated states.
PSI techniques are used to accurately measure the phase distribution of a wavefront, and they have proven useful for characterizing OAM modes \cite{Huang:13} and also for the
reconstruction of pure quantum states \cite{Stefano2017}. For this purpose, successive phase shifts are introduced in the reference beam that interferes with the object beam. In each step, the intensity distribution of the interferogram is given by $I(\boldsymbol{\rho};\Delta) = I_{0}+\gamma\cos{(\Phi(\boldsymbol{\rho})+\Delta)}$,
where $\Phi(\boldsymbol{\rho})$ represents the phase of interest, $\Delta$ is the constant phase value added to the reference in each step $\left(\Delta=0,\pi/2, \pi, 3\pi/2\right)$, $I_{0}$ is the arithmetic sum of the
intensity of the reference beam ($I_1$) plus the intensity of the object beam ($I_2$), and $\gamma=\frac{2\sqrt{I_{1}I_{2}}}{I_{0}}$ is the
modulation of the interference fringes. From here, the phase distribution is calculated as $\Phi(\boldsymbol{\rho}) = \mathrm{tan}^{-1}{\left(\frac{I(\boldsymbol{\rho};3\pi/2)-I(\boldsymbol{\rho};\pi/2)}{I(\boldsymbol{\rho};0)-I(\boldsymbol{\rho};\pi)}\right)}$.
In our experimental implementation,
the regions on the SLM external to the fork holograms do not contain any information about the OAM-transverse path state. Hence, this background region can be used for generating the reference beam for a full characterization of the state. 
Unlike Ref.~\cite{Andersen:19}, we have chosen a non-collinear interferometry.
As it is schematically shown in Fig.~\ref{fig_1}(a), we display a tilted blazed grating in the region without information, sending the reference beam to a different position in the far field to which the state information is sent. The required phase shifting $\Delta$ in each step of the PSI is achieved by a lateral displacement of this grating. 
\begin{figure}[h]
\centering
\includegraphics[width=0.47\textwidth]{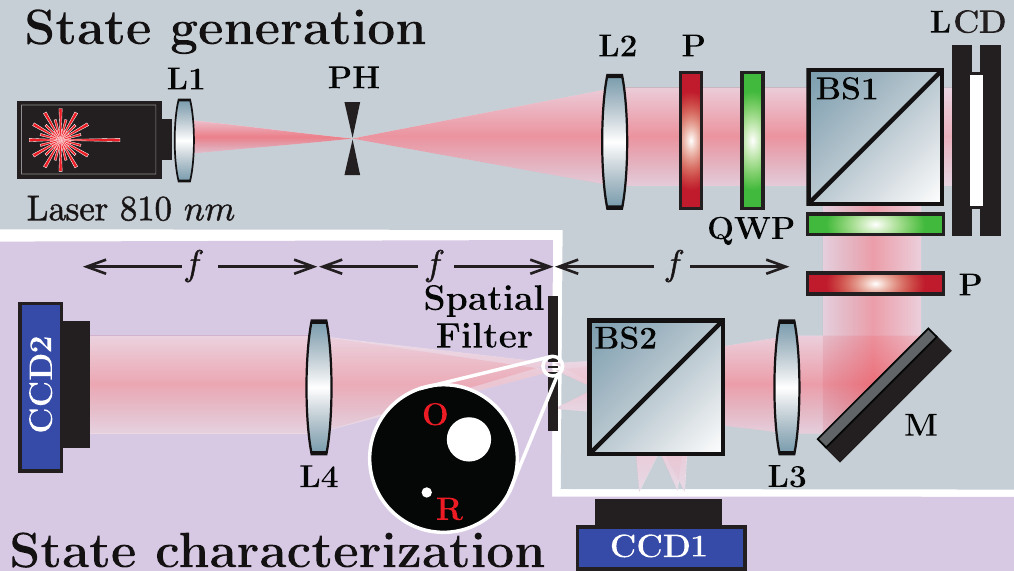}
\caption{Schematic experimental setup: Lens (L), pinhole (PH), beam splitter (BS), linear polarizer (P), quarter wave plate (QWP), liquid crystal display (LCD), mirror (M), charge-coupled device camera (CCD), object pinhole (O) and reference pinhole  (R). 
}\label{fig_setup}
\end{figure}

Figure~\ref{fig_setup} shows the experimental setup used for the generation and characterization of the states described by Eq.~(\ref{eq1}). In the first part, where the state is generated, a $810\,nm$ laser diode 
is expanded, filtered and collimated. 
Linear polarizers (P) and quarter wave plates (QWP) are set to control the polarization of light both at the input and output of a reflective LCD (Holoeye LC-R 2500). We used elliptical polarization since it allows us to modulate mostly the phase~\cite{Iemmi2001} of the wavefront as required by our encoding. The modulated light is split in two arms by the beam splitter BS2. On each arm the first diffracted order, which carries the required information, is selected. In the focal plane of the lens L3 we placed a spatial filter containing two pinholes: one for filtering the object beam (O) and the other one for selecting light as the reference beam (R). The pinhole sizes were chosen to allow the whole information to pass in the case of the object beam (1 mm of diameter for O), and small enough in the case of the reference beam so that it can be approximated to a point source (0.1 mm of diameter for R).
In the second part of the setup, the state characterization is performed. The intensity measurements are carried out with charge-coupled device (CCD) cameras. The far-field distribution is registered by CCD1 in one of the arms, while in the other arm the image of the LCD is obtained by CCD2.
It is important to remark that an implementation of the state preparation and reconstructing method by using a single photon source requires exactly the
same setup except for the use of high-sensitive cameras for single photon detection \cite{fickler2013real,Bolduc_2017}.
\begin{figure}[h]
\centering
\includegraphics[width=0.47\textwidth]{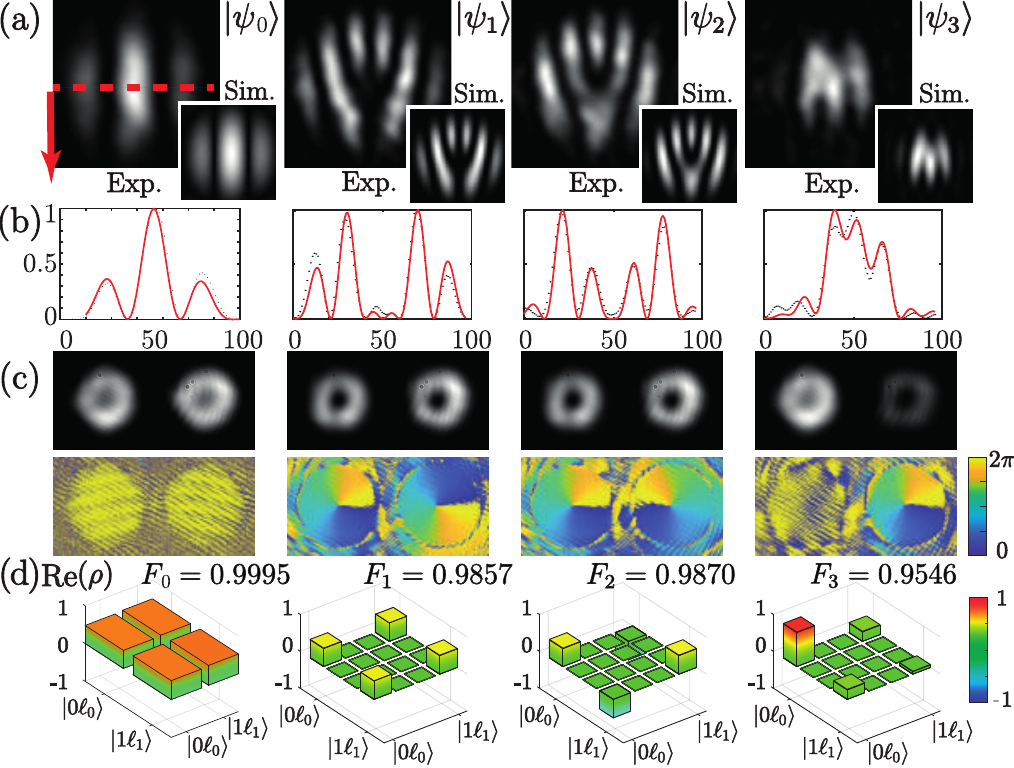}
\caption{Reconstruction of different two-qubit states. (a) Experimental and simulated far-field profiles. (b) Measured (black) and simulated (red) intensity of the one-dimensional profile obtained from (a). (c) Real amplitude (top) and phase profiles (bottom) registered in the near field. (d) Real part of the reconstructed density matrices $\rho_i$ where
$\{\ell_{0},\ell_{1}\}= \{0,0\}$ for $\ket{\psi_{0}}$, $\{1,-1\}$ for $\ket{\psi_{1,2}}$, and $\{0,-1\}$ for $\ket{\psi_{3}}$. The imaginary parts of $\rho_i$, not shown here, are null in agreement with the expected ones.
}\label{fig_3} 
\end{figure}

\begin{figure}[h]
\centering
\includegraphics[width=0.47\textwidth]{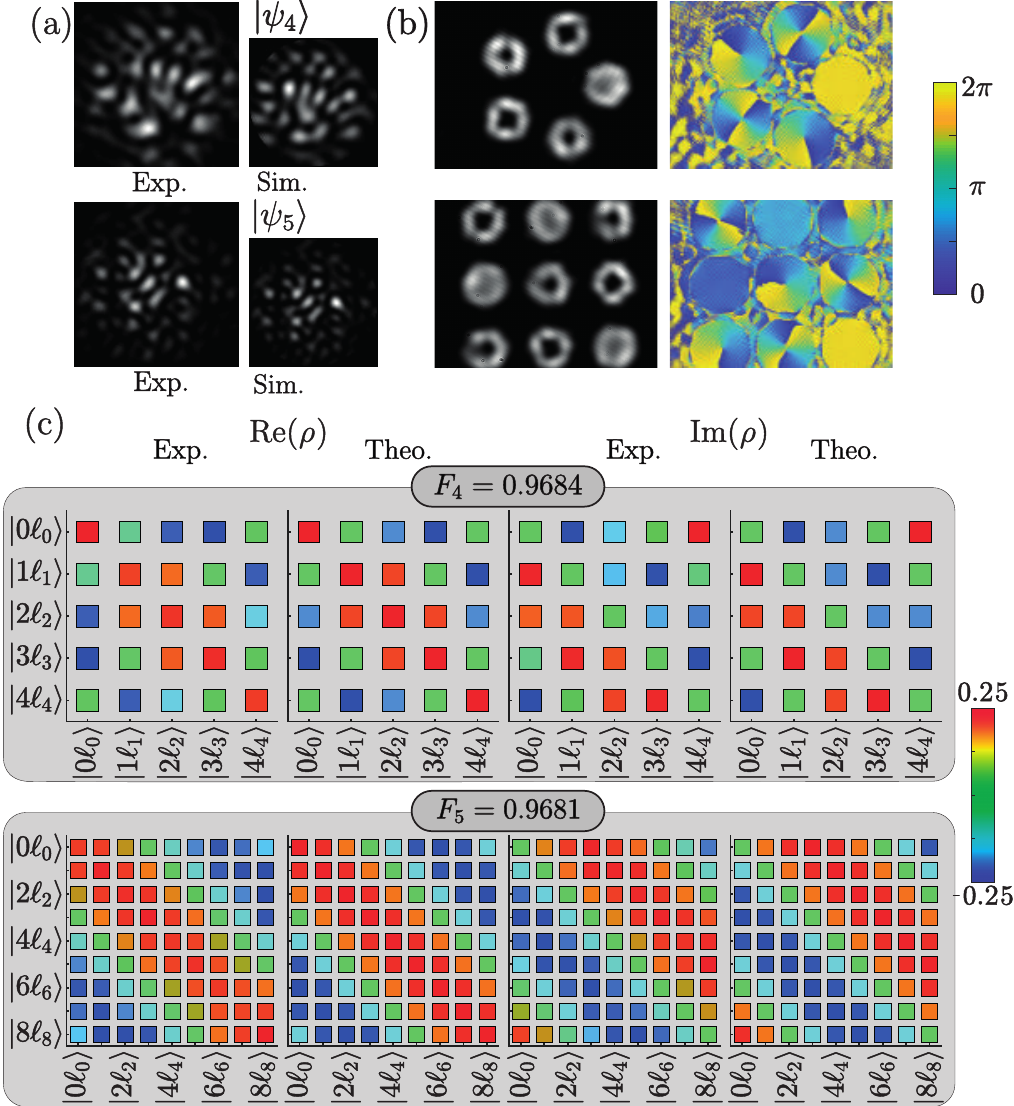}
\caption{Reconstruction of the states $\ket{\psi_{4}}$ and $\ket{\psi_{5}}$ which emulate a two-qudit system of dimensions $d=m=5$, and  $d=9$ and $m=5$, respectively. 
(a) Experimental and simulated far-field profiles. (b) Real amplitude (left) and phase profiles (right) registered in the near field. (c) Real part of the non-zero elements of the reconstructed density matrices $\rho_i$ in comparison with the theoretical ones where $\{\ell_{\mu}\}_{\mu=0}^{d-1} = \{0,1,-2,-1,2\}$ for $\ket{\psi_{4}}$ and $\{-2,0,1,0,-2,2,1,-2,0\}$ for $\ket{\psi_{5}}$.}
\label{fig_4}
\end{figure}

\textit{Results and discussion.} 
To test the feasibility of the proposed method for generating OAM-transverse path states we start by using an aperture $T(\boldsymbol{\rho})$ with $d=2$ like the one shown in the scheme of~Fig.\ref{fig_1}(a). In each of the $d$ circular regions defining the dimension of the path sector we displayed a binary fork hologram of 4 pixels of period. It was multiplied by a blazed grating of 10 pixels of period in the vertical direction. Each region has a diameter of 56 pixels with a separation between centers of 112 pixels. Outside the circular regions, a second blazed grating that generates the reference for the PSI was oriented at $190^{\circ}$ with respect to the previous ones. It is enough for sending the backlight in the far field to a region distant from the light coming from the fork holograms. As examples, the states $\ket{\psi_{0}}=\frac{\ket{00}+\ket{10}}{\sqrt{2}}$, $\ket{\psi_{1,2}}=\frac{\ket{0-1}\pm\ket{11}}{\sqrt{2}}$, 
and $\ket{\psi_{3}}=\frac{3\ket{00}+\ket{1-1}}{\sqrt{10}}$ were implemented. These states emulate different types of two-qubit states: 
a separable state $\ket{\psi_{0}}$, two maximally non-separable states $\ket{\psi_{1,2}}$, and a partial non-separable state $\ket{\psi_{3}}$.
Figure~\ref{fig_3}(a) shows the measured and simulated far-field distributions for these states. In Fig.~\ref{fig_3}(b) we show the one-dimensional profiles corresponding to the horizontal line trough the center of the images in Fig.~\ref{fig_3}(a) (indicated in the first image with a red dotted line).
One can clearly see the interference fringe structures corresponding to a coherent superposition of two beams with the given complex amplitudes.
Figure~\ref{fig_3}(c) shows the intensity distributions obtained in the near field after the spatial filtering and the corresponding phase distributions reconstructed by the PSI method. On the one hand, the intensity profile that we see here has the typical doughnut shape where some irregularities are due in part to missfocusing and in part to asymmetries of the pinhole in the spatial filtering. However, this does not appreciably alter the results of the intensity and phase measurements we use to reconstruct the states.
On the other hand, we can see the typical spiral phases of a vortex beam anticipating the topological charge of each transverse path. Finally, the density matrices that bring the full description of the states, are plotted in Fig.~\ref{fig_3}(d). The quality of preparation, i.e., the overlapping between the prepared and the target state, can be measured by the fidelity $F$ and is above $95\%$ for all cases.

In Fig.~\ref{fig_4} we show the results of the proposed encoding for five and nine circular regions represented onto the LCD in two different spatial arrays. As an example, two arbitrary states of dimensions $5\times 5$ and $9\times 5$ were represented: $\ket{\psi_{4}}=\frac{1}{\sqrt{5}}\sum^{4}_{\mu=0} {e^{i\pi(\mu+1)/3}\ket{\mu \ell_{\mu}}}$ and $\ket{\psi_{5}}=\frac{1}{\sqrt{9}}\sum^{8}_{\mu=0} {e^{i\pi(\mu+1)/6}\ket{\mu \ell_{\mu}}}$, with $\{\ell_{\mu}\}_{\mu=0}^{d-1} = \{0,1,-2,-1,2\}$ and $\{-2,0,1,0,-2,2,1,-2,0\}$, respectively. In these cases, each circular region onto the LCD has a $56$-pixels diameter while the centers are separated by $44$ pixels.
The Fig.~\ref{fig_4}(a) shows the experimental and numerical results for the generated state in the far field. The phase and amplitude distributions obtained from the measurement in the near field are shown in Fig.~\ref{fig_4}(b) and their corresponding density matrices in Fig.~\ref{fig_4}(c). Even for this high-dimensional states, fidelities are higher than $96\%$. 

It should be note that, in general, the fidelities of the reconstructed states will decrease both by the number $d$ of spatial regions and the maximum topological charge $\ell$ that can be represented. This is mainly due to the limitation of generating fork holograms with a low number of pixels. Thus, the performance of the proposal codification for a given state is limited, in our case, by the resolution of the LCD.

To summarize, we have implemented a method that exploits the versatility of programmable SLMs to control and correlate two spatial DoFs of light. It can be useful for encoding high-dimensional  OAM-transverse path states, increasing the ability of photonic systems to handle information. The method is based on the encoding of fork holograms using superimposed phase gratings and a spatial filtering process. We were able to generate photonic states of dimension $2\times2$, $5\times5$ and $9\times5$ with high fidelity of preparation (above $95\%$). The proposed compact setup exploits the benefits of using a binary phase grating in combination with blazed gratings to maximize light efficiency, reduce noise and to characterize the generated states by using the same SLM system.

\textbf{Funding:} This work was supported by ANPCyT-PICT 2014-2432, CONICET-PIP 112200801-03047 and UBACyT 20020170100564BA. D. Pab\'on was supported by a CONICET Fellowship.

\bibliography{Biblio}

\begin{thebibliography}{27}
\expandafter\ifx\csname natexlab\endcsname\relax\def\natexlab#1{#1}\fi
\expandafter\ifx\csname bibnamefont\endcsname\relax
  \def\bibnamefont#1{#1}\fi
\expandafter\ifx\csname bibfnamefont\endcsname\relax
  \def\bibfnamefont#1{#1}\fi
\expandafter\ifx\csname citenamefont\endcsname\relax
  \def\citenamefont#1{#1}\fi
\expandafter\ifx\csname url\endcsname\relax
  \def\url#1{\texttt{#1}}\fi
\expandafter\ifx\csname urlprefix\endcsname\relax\def\urlprefix{URL }\fi
\providecommand{\bibinfo}[2]{#2}
\providecommand{\eprint}[2][]{\url{#2}}

\bibitem[{\citenamefont{Kwiat et~al.}(1995)\citenamefont{Kwiat, Mattle,
  Weinfurter, Zeilinger, Sergienko, and Shih}}]{PhysRevLett754337}
\bibinfo{author}{\bibfnamefont{P.~G.} \bibnamefont{Kwiat}},
  \bibinfo{author}{\bibfnamefont{K.}~\bibnamefont{Mattle}},
  \bibinfo{author}{\bibfnamefont{H.}~\bibnamefont{Weinfurter}},
  \bibinfo{author}{\bibfnamefont{A.}~\bibnamefont{Zeilinger}},
  \bibinfo{author}{\bibfnamefont{A.~V.} \bibnamefont{Sergienko}},
  \bibnamefont{and} \bibinfo{author}{\bibfnamefont{Y.}~\bibnamefont{Shih}},
  \bibinfo{journal}{Phys. Rev. Lett.} \textbf{\bibinfo{volume}{75}},
  \bibinfo{pages}{4337} (\bibinfo{year}{1995}).

\bibitem[{\citenamefont{Lukens and Lougovski}(2017)}]{Lukens2017}
\bibinfo{author}{\bibfnamefont{J.}~\bibnamefont{Lukens}} \bibnamefont{and}
  \bibinfo{author}{\bibfnamefont{P.}~\bibnamefont{Lougovski}},
  \bibinfo{journal}{Optica} \textbf{\bibinfo{volume}{4}}, \bibinfo{pages}{8}
  (\bibinfo{year}{2017}).

\bibitem[{\citenamefont{Brecht et~al.}(2015)\citenamefont{Brecht, Reddy,
  Silberhorn, and Raymer}}]{Brecht2015}
\bibinfo{author}{\bibfnamefont{B.}~\bibnamefont{Brecht}},
  \bibinfo{author}{\bibfnamefont{D.~V.} \bibnamefont{Reddy}},
  \bibinfo{author}{\bibfnamefont{C.}~\bibnamefont{Silberhorn}},
  \bibnamefont{and} \bibinfo{author}{\bibfnamefont{M.~G.}
  \bibnamefont{Raymer}}, \bibinfo{journal}{Phys. Rev. X}
  \textbf{\bibinfo{volume}{5}}, \bibinfo{pages}{041017} (\bibinfo{year}{2015}).

\bibitem[{\citenamefont{Allen et~al.}(1992)\citenamefont{Allen, Beijersbergen,
  Spreeuw, and Woerdman}}]{Allen92}
\bibinfo{author}{\bibfnamefont{L.}~\bibnamefont{Allen}},
  \bibinfo{author}{\bibfnamefont{M.~W.} \bibnamefont{Beijersbergen}},
  \bibinfo{author}{\bibfnamefont{R.~J.~C.} \bibnamefont{Spreeuw}},
  \bibnamefont{and} \bibinfo{author}{\bibfnamefont{J.~P.}
  \bibnamefont{Woerdman}}, \bibinfo{journal}{Phys. Rev. A}
  \textbf{\bibinfo{volume}{45}}, \bibinfo{pages}{8185} (\bibinfo{year}{1992}).

\bibitem[{\citenamefont{Erhard et~al.}(2018)\citenamefont{Erhard, Fickler,
  Krenn, and Zeilinger}}]{Erhard2018}
\bibinfo{author}{\bibfnamefont{M.}~\bibnamefont{Erhard}},
  \bibinfo{author}{\bibfnamefont{R.}~\bibnamefont{Fickler}},
  \bibinfo{author}{\bibfnamefont{M.}~\bibnamefont{Krenn}}, \bibnamefont{and}
  \bibinfo{author}{\bibfnamefont{A.}~\bibnamefont{Zeilinger}},
  \bibinfo{journal}{Light: Science and Applications}
  \textbf{\bibinfo{volume}{7}}, \bibinfo{pages}{17111} (\bibinfo{year}{2018}),
  ISSN \bibinfo{issn}{20477538}.

\bibitem[{\citenamefont{Krenn et~al.}(2016)\citenamefont{Krenn, Handsteiner,
  Fink, Fickler, Ursin, Malik, and Zeilinger}}]{Krenn13648}
\bibinfo{author}{\bibfnamefont{M.}~\bibnamefont{Krenn}},
  \bibinfo{author}{\bibfnamefont{J.}~\bibnamefont{Handsteiner}},
  \bibinfo{author}{\bibfnamefont{M.}~\bibnamefont{Fink}},
  \bibinfo{author}{\bibfnamefont{R.}~\bibnamefont{Fickler}},
  \bibinfo{author}{\bibfnamefont{R.}~\bibnamefont{Ursin}},
  \bibinfo{author}{\bibfnamefont{M.}~\bibnamefont{Malik}}, \bibnamefont{and}
  \bibinfo{author}{\bibfnamefont{A.}~\bibnamefont{Zeilinger}},
  \bibinfo{journal}{Proceedings of the National Academy of Sciences}
  \textbf{\bibinfo{volume}{113}}, \bibinfo{pages}{13648}
  (\bibinfo{year}{2016}), ISSN \bibinfo{issn}{0027-8424}.

\bibitem[{\citenamefont{Zhu et~al.}(2018)\citenamefont{Zhu, Wang, Chen, Liu,
  and Wang}}]{Zhu2018}
\bibinfo{author}{\bibfnamefont{L.}~\bibnamefont{Zhu}},
  \bibinfo{author}{\bibfnamefont{A.}~\bibnamefont{Wang}},
  \bibinfo{author}{\bibfnamefont{S.}~\bibnamefont{Chen}},
  \bibinfo{author}{\bibfnamefont{J.}~\bibnamefont{Liu}}, \bibnamefont{and}
  \bibinfo{author}{\bibfnamefont{J.}~\bibnamefont{Wang}},
  \bibinfo{journal}{Opt. Lett.} \textbf{\bibinfo{volume}{43}},
  \bibinfo{pages}{1894} (\bibinfo{year}{2018}).

\bibitem[{\citenamefont{Mirhosseini et~al.}(2015)\citenamefont{Mirhosseini,
  Maga{\~{n}}a-Loaiza, O'Sullivan, Rodenburg, Malik, Lavery, Padgett, Gauthier,
  and Boyd}}]{Mirhosseini_2015}
\bibinfo{author}{\bibfnamefont{M.}~\bibnamefont{Mirhosseini}},
  \bibinfo{author}{\bibfnamefont{O.~S.} \bibnamefont{Maga{\~{n}}a-Loaiza}},
  \bibinfo{author}{\bibfnamefont{M.~N.} \bibnamefont{O'Sullivan}},
  \bibinfo{author}{\bibfnamefont{B.}~\bibnamefont{Rodenburg}},
  \bibinfo{author}{\bibfnamefont{M.}~\bibnamefont{Malik}},
  \bibinfo{author}{\bibfnamefont{M.~P.~J.} \bibnamefont{Lavery}},
  \bibinfo{author}{\bibfnamefont{M.~J.} \bibnamefont{Padgett}},
  \bibinfo{author}{\bibfnamefont{D.~J.} \bibnamefont{Gauthier}},
  \bibnamefont{and} \bibinfo{author}{\bibfnamefont{R.~W.} \bibnamefont{Boyd}},
  \bibinfo{journal}{New Journal of Physics} \textbf{\bibinfo{volume}{17}},
  \bibinfo{pages}{033033} (\bibinfo{year}{2015}),
  \urlprefix\url{https://doi.org/10.1088%2F1367-2630%2F17%2F3%2F033033}.

\bibitem[{\citenamefont{He et~al.}(2017)\citenamefont{He, Zhu, Hu, and
  Jing}}]{He2017}
\bibinfo{author}{\bibfnamefont{G.}~\bibnamefont{He}},
  \bibinfo{author}{\bibfnamefont{C.}~\bibnamefont{Zhu}},
  \bibinfo{author}{\bibfnamefont{L.}~\bibnamefont{Hu}}, \bibnamefont{and}
  \bibinfo{author}{\bibfnamefont{J.}~\bibnamefont{Jing}},
  \bibinfo{journal}{Optics InfoBase Conference Papers}
  \textbf{\bibinfo{volume}{Part F54-NLO 2017}}, \bibinfo{pages}{1}
  (\bibinfo{year}{2017}).

\bibitem[{\citenamefont{Wang et~al.}(2018)\citenamefont{Wang, Luo, Huang, Chen,
  Su, Liu, Chen, Li, Fang, Jiang et~al.}}]{Wang2018}
\bibinfo{author}{\bibfnamefont{X.-L.} \bibnamefont{Wang}},
  \bibinfo{author}{\bibfnamefont{Y.}~\bibnamefont{Luo}},
  \bibinfo{author}{\bibfnamefont{H.-L.} \bibnamefont{Huang}},
  \bibinfo{author}{\bibfnamefont{M.-c.} \bibnamefont{Chen}},
  \bibinfo{author}{\bibfnamefont{Z.-E.} \bibnamefont{Su}},
  \bibinfo{author}{\bibfnamefont{C.}~\bibnamefont{Liu}},
  \bibinfo{author}{\bibfnamefont{C.}~\bibnamefont{Chen}},
  \bibinfo{author}{\bibfnamefont{W.}~\bibnamefont{Li}},
  \bibinfo{author}{\bibfnamefont{Y.-Q.} \bibnamefont{Fang}},
  \bibinfo{author}{\bibfnamefont{X.}~\bibnamefont{Jiang}},
  \bibnamefont{et~al.}, \bibinfo{journal}{Physical Review Letters}
  \textbf{\bibinfo{volume}{120}} (\bibinfo{year}{2018}).

\bibitem[{\citenamefont{Hashemi~Rafsanjani
  et~al.}(2015)\citenamefont{Hashemi~Rafsanjani, Mirhosseini, Maga\~na Loaiza,
  and Boyd}}]{Rafsanjani2015}
\bibinfo{author}{\bibfnamefont{S.~M.} \bibnamefont{Hashemi~Rafsanjani}},
  \bibinfo{author}{\bibfnamefont{M.}~\bibnamefont{Mirhosseini}},
  \bibinfo{author}{\bibfnamefont{O.~S.} \bibnamefont{Maga\~na Loaiza}},
  \bibnamefont{and} \bibinfo{author}{\bibfnamefont{R.~W.} \bibnamefont{Boyd}},
  \bibinfo{journal}{Phys. Rev. A} \textbf{\bibinfo{volume}{92}},
  \bibinfo{pages}{023827} (\bibinfo{year}{2015}),
  \urlprefix\url{https://link.aps.org/doi/10.1103/PhysRevA.92.023827}.

\bibitem[{\citenamefont{Aiello et~al.}(2015)\citenamefont{Aiello, Töppel,
  Marquardt, Giacobino, and Leuchs}}]{Aiello2015}
\bibinfo{author}{\bibfnamefont{A.}~\bibnamefont{Aiello}},
  \bibinfo{author}{\bibfnamefont{F.}~\bibnamefont{Töppel}},
  \bibinfo{author}{\bibfnamefont{C.}~\bibnamefont{Marquardt}},
  \bibinfo{author}{\bibfnamefont{E.}~\bibnamefont{Giacobino}},
  \bibnamefont{and} \bibinfo{author}{\bibfnamefont{G.}~\bibnamefont{Leuchs}},
  \bibinfo{journal}{New Journal of Physics} \textbf{\bibinfo{volume}{17}},
  \bibinfo{pages}{043024} (\bibinfo{year}{2015}).

\bibitem[{\citenamefont{Ndagano et~al.}(2017)\citenamefont{Ndagano,
  Perez-Garcia, Roux, McLaren, Rosales-Guzman, Zhang, Mouane, Hernandez-Aranda,
  Konrad, and Forbes}}]{ndagano2017characterizing}
\bibinfo{author}{\bibfnamefont{B.}~\bibnamefont{Ndagano}},
  \bibinfo{author}{\bibfnamefont{B.}~\bibnamefont{Perez-Garcia}},
  \bibinfo{author}{\bibfnamefont{F.~S.} \bibnamefont{Roux}},
  \bibinfo{author}{\bibfnamefont{M.}~\bibnamefont{McLaren}},
  \bibinfo{author}{\bibfnamefont{C.}~\bibnamefont{Rosales-Guzman}},
  \bibinfo{author}{\bibfnamefont{Y.}~\bibnamefont{Zhang}},
  \bibinfo{author}{\bibfnamefont{O.}~\bibnamefont{Mouane}},
  \bibinfo{author}{\bibfnamefont{R.~I.} \bibnamefont{Hernandez-Aranda}},
  \bibinfo{author}{\bibfnamefont{T.}~\bibnamefont{Konrad}}, \bibnamefont{and}
  \bibinfo{author}{\bibfnamefont{A.}~\bibnamefont{Forbes}},
  \bibinfo{journal}{Nature Physics} \textbf{\bibinfo{volume}{13}},
  \bibinfo{pages}{397} (\bibinfo{year}{2017}).

\bibitem[{\citenamefont{Töppel et~al.}(2014)\citenamefont{Töppel, Aiello,
  Marquardt, Giacobino, and Leuchs}}]{Tppel2014}
\bibinfo{author}{\bibfnamefont{F.}~\bibnamefont{Töppel}},
  \bibinfo{author}{\bibfnamefont{A.}~\bibnamefont{Aiello}},
  \bibinfo{author}{\bibfnamefont{C.}~\bibnamefont{Marquardt}},
  \bibinfo{author}{\bibfnamefont{E.}~\bibnamefont{Giacobino}},
  \bibnamefont{and} \bibinfo{author}{\bibfnamefont{G.}~\bibnamefont{Leuchs}},
  \bibinfo{journal}{New Journal of Physics} \textbf{\bibinfo{volume}{16}},
  \bibinfo{pages}{073019} (\bibinfo{year}{2014}).

\bibitem[{\citenamefont{Rosales-Guzm{\'{a}}n and
  Forbes}(2017)}]{RosalesGuzman2017}
\bibinfo{author}{\bibfnamefont{C.}~\bibnamefont{Rosales-Guzm{\'{a}}n}}
  \bibnamefont{and} \bibinfo{author}{\bibfnamefont{A.}~\bibnamefont{Forbes}},
  \emph{\bibinfo{title}{{How to Shape Light with Spatial Light Modulators}}}
  (\bibinfo{publisher}{SPIE PRESS}, \bibinfo{year}{2017}).

\bibitem[{\citenamefont{Lima et~al.}(2009)\citenamefont{Lima, Vargas, Neves,
  Guzm\'{a}n, and Saavedra}}]{Lima2009}
\bibinfo{author}{\bibfnamefont{G.}~\bibnamefont{Lima}},
  \bibinfo{author}{\bibfnamefont{A.}~\bibnamefont{Vargas}},
  \bibinfo{author}{\bibfnamefont{L.}~\bibnamefont{Neves}},
  \bibinfo{author}{\bibfnamefont{R.}~\bibnamefont{Guzm\'{a}n}},
  \bibnamefont{and} \bibinfo{author}{\bibfnamefont{C.}~\bibnamefont{Saavedra}},
  \bibinfo{journal}{Opt. Express} \textbf{\bibinfo{volume}{17}},
  \bibinfo{pages}{10688} (\bibinfo{year}{2009}).

\bibitem[{\citenamefont{Sol{\'i}s-Prosser
  et~al.}(2013)\citenamefont{Sol{\'i}s-Prosser, Arias, Varga, Reb{\'o}n,
  Ledesma, Iemmi, and Neves}}]{SolsProsser2013PreparingAP}
\bibinfo{author}{\bibfnamefont{M.~A.} \bibnamefont{Sol{\'i}s-Prosser}},
  \bibinfo{author}{\bibfnamefont{A.}~\bibnamefont{Arias}},
  \bibinfo{author}{\bibfnamefont{J.~J.} \bibnamefont{Varga}},
  \bibinfo{author}{\bibfnamefont{L.}~\bibnamefont{Reb{\'o}n}},
  \bibinfo{author}{\bibfnamefont{S.~G.} \bibnamefont{Ledesma}},
  \bibinfo{author}{\bibfnamefont{C.~C.} \bibnamefont{Iemmi}}, \bibnamefont{and}
  \bibinfo{author}{\bibfnamefont{L.}~\bibnamefont{Neves}},
  \bibinfo{journal}{Optics letters} \textbf{\bibinfo{volume}{38 22}},
  \bibinfo{pages}{4762} (\bibinfo{year}{2013}).

\bibitem[{\citenamefont{Varga et~al.}(2014)\citenamefont{Varga, Reb{\'{o}}n,
  Sol{\'{\i}}s-Prosser, Neves, Ledesma, and Iemmi}}]{Varga2014}
\bibinfo{author}{\bibfnamefont{J.~J.~M.} \bibnamefont{Varga}},
  \bibinfo{author}{\bibfnamefont{L.}~\bibnamefont{Reb{\'{o}}n}},
  \bibinfo{author}{\bibfnamefont{M.~A.} \bibnamefont{Sol{\'{\i}}s-Prosser}},
  \bibinfo{author}{\bibfnamefont{L.}~\bibnamefont{Neves}},
  \bibinfo{author}{\bibfnamefont{S.}~\bibnamefont{Ledesma}}, \bibnamefont{and}
  \bibinfo{author}{\bibfnamefont{C.}~\bibnamefont{Iemmi}},
  \bibinfo{journal}{Journal of Physics B: Atomic, Molecular and Optical
  Physics} \textbf{\bibinfo{volume}{47}}, \bibinfo{pages}{225504}
  (\bibinfo{year}{2014}).

\bibitem[{\citenamefont{Creath}(1988)}]{Creath1988}
\bibinfo{author}{\bibfnamefont{K.}~\bibnamefont{Creath}},
  \bibinfo{journal}{Progress in Optics} \textbf{\bibinfo{volume}{26}},
  \bibinfo{pages}{349} (\bibinfo{year}{1988}), ISSN \bibinfo{issn}{00796638}.

\bibitem[{\citenamefont{Pabon et~al.}(2017)\citenamefont{Pabon, A.~Ledesma,
  Quinteiro, and Capeluto}}]{Pabon20172}
\bibinfo{author}{\bibfnamefont{D.}~\bibnamefont{Pabon}},
  \bibinfo{author}{\bibfnamefont{S.}~\bibnamefont{A.~Ledesma}},
  \bibinfo{author}{\bibfnamefont{G.}~\bibnamefont{Quinteiro}},
  \bibnamefont{and} \bibinfo{author}{\bibfnamefont{M.}~\bibnamefont{Capeluto}},
  \bibinfo{journal}{Applied Optics} \textbf{\bibinfo{volume}{56}},
  \bibinfo{pages}{8048} (\bibinfo{year}{2017}).

\bibitem[{\citenamefont{Goodman}(2005)}]{goodman2005introduction}
\bibinfo{author}{\bibfnamefont{J.~W.} \bibnamefont{Goodman}},
  \emph{\bibinfo{title}{Introduction to Fourier optics}}
  (\bibinfo{publisher}{Roberts and Company Publishers}, \bibinfo{year}{2005}).

\bibitem[{\citenamefont{Huang et~al.}(2013)\citenamefont{Huang, Ren, Yan,
  Ahmed, Yue, Bozovich, Erkmen, Birnbaum, Dolinar, Tur et~al.}}]{Huang:13}
\bibinfo{author}{\bibfnamefont{H.}~\bibnamefont{Huang}},
  \bibinfo{author}{\bibfnamefont{Y.}~\bibnamefont{Ren}},
  \bibinfo{author}{\bibfnamefont{Y.}~\bibnamefont{Yan}},
  \bibinfo{author}{\bibfnamefont{N.}~\bibnamefont{Ahmed}},
  \bibinfo{author}{\bibfnamefont{Y.}~\bibnamefont{Yue}},
  \bibinfo{author}{\bibfnamefont{A.}~\bibnamefont{Bozovich}},
  \bibinfo{author}{\bibfnamefont{B.~I.} \bibnamefont{Erkmen}},
  \bibinfo{author}{\bibfnamefont{K.}~\bibnamefont{Birnbaum}},
  \bibinfo{author}{\bibfnamefont{S.}~\bibnamefont{Dolinar}},
  \bibinfo{author}{\bibfnamefont{M.}~\bibnamefont{Tur}}, \bibnamefont{et~al.},
  \bibinfo{journal}{Opt. Lett.} \textbf{\bibinfo{volume}{38}},
  \bibinfo{pages}{2348} (\bibinfo{year}{2013}),
  \urlprefix\url{http://ol.osa.org/abstract.cfm?URI=ol-38-13-2348}.

\bibitem[{\citenamefont{Pears~Stefano et~al.}(2017)\citenamefont{Pears~Stefano,
  Reb\'on, Ledesma, and Iemmi}}]{Stefano2017}
\bibinfo{author}{\bibfnamefont{Q.}~\bibnamefont{Pears~Stefano}},
  \bibinfo{author}{\bibfnamefont{L.}~\bibnamefont{Reb\'on}},
  \bibinfo{author}{\bibfnamefont{S.}~\bibnamefont{Ledesma}}, \bibnamefont{and}
  \bibinfo{author}{\bibfnamefont{C.}~\bibnamefont{Iemmi}},
  \bibinfo{journal}{Phys. Rev. A} \textbf{\bibinfo{volume}{96}},
  \bibinfo{pages}{062328} (\bibinfo{year}{2017}),
  \urlprefix\url{https://link.aps.org/doi/10.1103/PhysRevA.96.062328}.

\bibitem[{\citenamefont{Andersen et~al.}(2019)\citenamefont{Andersen, Alperin,
  Voitiv, Holtzmann, Gopinath, and Siemens}}]{Andersen:19}
\bibinfo{author}{\bibfnamefont{J.~M.} \bibnamefont{Andersen}},
  \bibinfo{author}{\bibfnamefont{S.~N.} \bibnamefont{Alperin}},
  \bibinfo{author}{\bibfnamefont{A.~A.} \bibnamefont{Voitiv}},
  \bibinfo{author}{\bibfnamefont{W.~G.} \bibnamefont{Holtzmann}},
  \bibinfo{author}{\bibfnamefont{J.~T.} \bibnamefont{Gopinath}},
  \bibnamefont{and} \bibinfo{author}{\bibfnamefont{M.~E.}
  \bibnamefont{Siemens}}, \bibinfo{journal}{Appl. Opt.}
  \textbf{\bibinfo{volume}{58}}, \bibinfo{pages}{404} (\bibinfo{year}{2019}),
  \urlprefix\url{http://ao.osa.org/abstract.cfm?URI=ao-58-2-404}.

\bibitem[{\citenamefont{Marquez et~al.}(2001)\citenamefont{Marquez, Iemmi,
  Moreno, Davis, Campos, and Yzuel}}]{Iemmi2001}
\bibinfo{author}{\bibfnamefont{A.}~\bibnamefont{Marquez}},
  \bibinfo{author}{\bibfnamefont{C.~C.} \bibnamefont{Iemmi}},
  \bibinfo{author}{\bibfnamefont{I.~S.} \bibnamefont{Moreno}},
  \bibinfo{author}{\bibfnamefont{J.~A.} \bibnamefont{Davis}},
  \bibinfo{author}{\bibfnamefont{J.}~\bibnamefont{Campos}}, \bibnamefont{and}
  \bibinfo{author}{\bibfnamefont{M.~J.} \bibnamefont{Yzuel}},
  \bibinfo{journal}{Optical Engineering} \textbf{\bibinfo{volume}{40}},
  \bibinfo{pages}{2558 } (\bibinfo{year}{2001}),
  \urlprefix\url{https://doi.org/10.1117/1.1412228}.

\bibitem[{\citenamefont{Fickler et~al.}(2013)\citenamefont{Fickler, Krenn,
  Lapkiewicz, Ramelow, and Zeilinger}}]{fickler2013real}
\bibinfo{author}{\bibfnamefont{R.}~\bibnamefont{Fickler}},
  \bibinfo{author}{\bibfnamefont{M.}~\bibnamefont{Krenn}},
  \bibinfo{author}{\bibfnamefont{R.}~\bibnamefont{Lapkiewicz}},
  \bibinfo{author}{\bibfnamefont{S.}~\bibnamefont{Ramelow}}, \bibnamefont{and}
  \bibinfo{author}{\bibfnamefont{A.}~\bibnamefont{Zeilinger}},
  \bibinfo{journal}{Scientific reports} \textbf{\bibinfo{volume}{3}}
  (\bibinfo{year}{2013}), \urlprefix\url{https://dx.doi.org/10.1038/srep01914}.

\bibitem[{\citenamefont{Bolduc et~al.}(2017)\citenamefont{Bolduc, Faccio, and
  Leach}}]{Bolduc_2017}
\bibinfo{author}{\bibfnamefont{E.}~\bibnamefont{Bolduc}},
  \bibinfo{author}{\bibfnamefont{D.}~\bibnamefont{Faccio}}, \bibnamefont{and}
  \bibinfo{author}{\bibfnamefont{J.}~\bibnamefont{Leach}},
  \bibinfo{journal}{Journal of Optics} \textbf{\bibinfo{volume}{19}},
  \bibinfo{pages}{054006} (\bibinfo{year}{2017}),
  \urlprefix\url{https://doi.org/10.1088%2F2040-8986%2Faa52d8}.

\end{thebibliography}
\end{document}